# Incremental extraction of a NoSQL database model using an MDA-based process

**Amal AIT BRAHIM, Rabah TIGHILT FERHAT and Gilles ZURFLUH**
Toulouse Institute of Computer Science Research (IRIT),
Toulouse Capitole University, Toulouse, France

**Abstract -** *In recent years, the need to use NoSQL systems to store and exploit big data has been steadily increasing. Most of these systems are characterized by the property "schema less" which means absence of the data model when creating a database. This property brings an undeniable flexibility by allowing the evolution of the model during the exploitation of the base. However, the expression of queries requires a precise knowledge of this model. In this paper, we propose an incremental process to extract the model while operating the document-oriented NoSQL database. To do this, we use the Model Driven Architecture (MDA) that provides a formal framework for automatic model transformation. From the insert, delete and update queries executed on the database, we propose formal transformation rules with QVT to generate the physical model of the NoSQL database. An experimentation of the extraction process was performed on a medical application.*

## 1 Introduction

The number of digital devices that we use nowadays produces a huge amount of data, known as Big Data that need to be exploited. Usually, who says Big Data, says at least, Volume, Variety and Velocity [5]. Volume is the size of the data set that needs to be processed, Variety describes different data type including factors such as format, structure, and sources and Velocity indicates the speed of data loading and processing. Relational systems that had been for decades the one solution for all databases needs prove to be inadequate for all applications, especially those involving Big Data [1]. Consequently, new type of DBMS, commonly known as “NoSQL” [8], has appeared. These systems are well suited for managing large volume of data with flexible models. They also provide low latency at scale and faster data access [1]. NoSQL covers a wide variety of different systems that were developed to meet specific needs. These systems can be classified into four basic types: key-value, column-oriented, document-oriented and graph-oriented. In this paper, we focus on the third one.

One of the NoSQL key features is that databases can be schema-less. This means, in a table, meanwhile the row is inserted, the attributes names and types are specified. The schema-less property offers undeniable flexibility that facilitates the database model evolution. End-users are able to quickly and easily incorporate new data into their applications without rewriting tables. However, the importance and the necessity of the database model are widely recognized. There is still a need for this model to know how data is structured and related in the database; this is particularly necessary to write declarative queries where tables and columns names are specified.

On the one hand, NoSQL systems have proven their efficiency to handle Big Data. On the other hand, the needs of the NoSQL database model remain up-to-date. Therefore, we are convinced that it’s important to provide an automatic approach that extracts the database model within NoSQL systems. To formalize and automate this approach, we have used the Model Driven Architecture (MDA) that is well known as a framework for models automatic transformations. Our approach starts from the user queries and update the database model. For each query provided by the user and applied to the database, the process has mainly two steps: (1) it analyzes the query and (2) update the database model by applying the changes introduced by this user query.

The remainder of the paper is structured as follows. Section 2 motivates our work using a case of study in the healthcare field. Section 3 reviews previous work. Section 4 defines our NoSQL database model extraction process. Section 5 details our experiments as well as the validation of our process. Finally, Section 6 concludes the paper and announces future work.

## 2 Motivation

To motivate and illustrate our work, we have used a case study in the healthcare filed. This case study concerns international scientific programs for monitoring patients suffering from serious diseases. The main goal of this program is (1) to collect data about diseases development over time, (2) to study interactions between different diseases and (3) to evaluate the short and medium-term effects of their treatments. The medical program can last up to 3 years. Data collected from establishments involved in this kind of program have the features of Big Data (the 3 V):
**Volume**: the amount of data collected from all the establishments in three years can reach several terabytes. **Variety**: data created while monitoring patients come in different types; it could be (1) structured as the patient's vital signs (respiratory rate, blood pressure, etc.), (2) semi-structured document such as the package leaflets of medicinal products, (3) unstructured such as consultation summaries, paper prescriptions and radiology reports. Velocity: some data are produced in continuous way by sensors; it needs a [near] real time process because it could be integrated into a time-sensitive processes (for example, some measurements, like temperature, require an emergency medical treatment if they cross a given threshold).

This is a typical example in which the use of a NoSQL system is suitable. On the one hand, in the medical application, briefly presented above, the database contains structured data, data of various types and formats (explanatory texts, medical records, x-rays, etc.), and big tables (records of variables produced by sensors). On the other hand, NoSQL data stores are ideally suited for this kind of applications that use large amounts of disparate data. Therefore, we are convinced that a NoSQL DBMS, like MongoDB, is the most adapted system to store the medical database.

As mentioned before, this kind of systems operate on schema-less data model. Nevertheless, there is still a need for the database model in order to know how data is structured and related in the database and then to express queries. Regarding the medical application, doctors enter measures regularly for a cohort of patients. They can also recording new data in cases where the patient's state of health evolve over time. Few months later, they will analyze the entered data in order to follow the evolution of the pathology. For this, they need the database model to express their queries.

In our view, it’s important to have a precise and automatic solution that guides and facilitates the database model extraction task within NoSQL systems. For this, we propose the Query2Model process presented in the next section that extracts the physical model of a database stored in MongoDB. This model is expressed using the JSON format.

# 3 Related work

Several research works have been proposed to extract a NoSQL databases model, mainly for document-oriented databases such as MongoDB. In [11], the authors present a process to extract a model from a collection of JSON documents stored on MongoDB. The model returned by this process is in JSON format; it is obtained by capturing the names of the attributes that appear in the input documents and replacing their values with their types. Attribute values can be atomic, lists, or nested documents.

Authors in [12] propose a model extraction process from a document-oriented NoSQL database that can include several collections. The returned result is not a unified model for the whole database but it is a set of model versions. These versions are stored in JSON format.

More specific to document-oriented databases, we can mention [7] where authors describe a process called BSP (Build Schema Profile) to classify the documents of a collection by applying a set of rules that correspond to the user requirements. These rules are expressed through a decision tree where nodes represent the attributes of the documents and edges specify the conditions on which the classification is based. These conditions reflect either the absence or the presence of an attribute in a document or its value. As in the previous article [12], the result returned by this approach is not a unified model but a set of model versions ; each of them is common to a group of documents.

We can also mention [14] that describes a mapping from a document-oriented NoSQL database to a relational model. The process groups together all documents that have the same fields name. For each class of documents, it generates a table that have as columns the fields names and as rows the fields values.

Another study [15] have proposed a model extraction process from a collection of JSON documents. This process is based on the use of MapReduce. The Map step consists of extracting the schema of each document in the collection by mapping each couple (field, value) into another couple (field, type). The Reduce step consists of unifying all the schemas produced in the Map step in order to provide an overall schema for the input collection. The same authors have proposed in another paper [16] an extension of the process prposed in [15] in order to take into account the parameterization of the extraction at the Reduce step. Thus, the user can choose either to unify all the schemas of the collection, or to unify only the schemas having the same fields ( same names and types).

On the other hand, [13] proposes a process for extracting a model from object insertion queries and relations in a graph-oriented databases. The proposed process is based on an MDA architecture and applies two treatments. The first one build a graph (Nodes + Edges) starting from Neo4j queries. The second one consists of extracting an Entity / Association model from the graph returned by the first treatment.

In Table 1, we summarize the previous works using three criteria: the database content (one or several classes), the considered NoSQL system type (document or graph) and the way used to implement links (references, nested data or edges).

| | Database content | | NoSQL system type | | Links | | | | | |
|---|---|---|---|---|---|---|---|---|---|---|
| | | | | | Intra classe | | | Inter classes | | |
| | One class | Several classes | Document-oriented | Graph-oriented | Using references | Using nested data | Using edges | Using references | Using nested data | Using edges |
| [11]: Klettke et al., 2015 | X | | X | | | X | | | | |
| [12]: Sevilla et al., 2015 | | X | X | | | X | | | X | |
| [7]: Gallinucci et al., 2018 | X | | X | | | X | | | | |
| [13]: Comyn-Wattiau et al., 2018 | | X | | X | | | X | | | X |
| [14]: Maity et al., 2018 | X | | X | | | X | | | | |
| [15]: Baazizi et al., 2017 | X | | X | | | X | | | | |
| [16]: Baazizi et al., 2019 | X | | X | | | X | | | | |

Table. 1 Comparative table of previous works

Regarding the state of the art, the solutions proposed in [7], [11], [14], [15] and [16] start from a single collection of documents and take into account only the links implemented using nested data ; the links presented using references are not considered. The process proposed in [12] takes as input a set of collections ; however, only the use of nested data to express links is considered. On the other hand, authors in [13] have worked on graph-oriented systems. This kind of NoSQL systems does not offer many solutions to implement links as like document-oriented systems ; it expresses explicitly links between data using edges. To overcome these limits, we define an automatic process to extract the database model within documents-oriented NoSQL systems. This process process takes into account the links between collections.

# 4 Query2Model process

This article focuses on extracting the model from a NoSQL database with the "schema less" property. We limit ourselves to the document-oriented type which is the most complete in terms of expression of links (use of references and nesting). For this, we propose the Query2Model process which automatically builds the model of a NoSQL database from update requests submitted by users.

The Query2Model process is based on OMG's Model Driven Architecture [10]. We recall below the outlines of this model transformation approach. MDA is a formal framework for formalizing and automating model transformations. The purpose of this architecture is to describe separately the functional specifications and implementation specifications of an application on a given platform. For this, MDA uses three models representing the abstraction levels of the application. These are (1) the Computational Independent Model (CIM) describing the services that the application must provide to meet the needs of users, (2) the analysis and design model (PIM for Platform Independent Model) which defines the structure and the behavior of the system without indicating the execution platform and (3) the model of code (PSM for Platform Specific Model) which is the projection of a PIM on a particular technical platform. Since the input of our process corresponds to user requests and its output is a physical model, we retain only the PSM level. The extraction of the model from a NoSQL database is done via a sequence of transformations. We will formalize these transformations using the QVT standard (Query View Transformation) defined by the OMG (see Experimentation). Figure 1 shows an overview of our process.

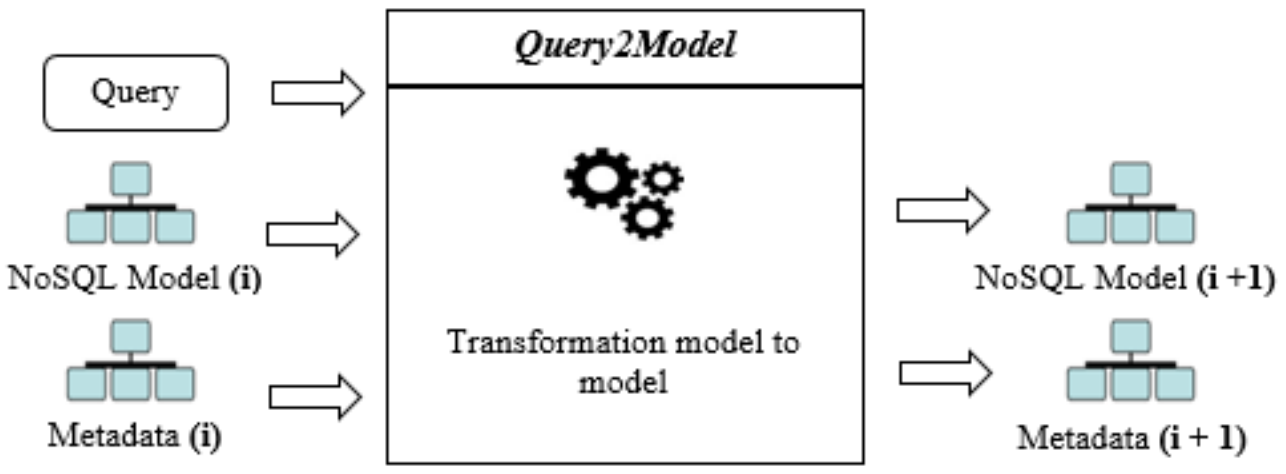

Fig. 1. Overview of Query2Model process

As shown in Figure 1, the input of the Query2Model process consists of three elements: (1) an insert, delete, or modify query, (2) a NoSQL database model, and (3) metadata. Note that at the first execution of the Query2Model process, the NoSQL model and the metadata are empty; these will be created and updated as requests are processed. Thus, for each new update request, the process proceeds as follows: (1) It analyzes the query against the model and available metadata, (2) It updates the model and enriches the metadata to allow the processing of future requests.

In the following sections, we detail the components of Query2Model by specifying the inputs / outputs as well as the transformation rules.

## 4.1 Inputs/ Outputs

### 4.1.1 Insertion query

An insertion query $Q^I$ is defined as a pair $(Cll, FL)$ where:

- $Q^I.Cll$ is the name of the collection on which $Q^I$ is executed,
- $Q^I.FL = AFL \cup CFL$ is the set of fields that appear on $Q^I$, where:
  - $AFL = \{afl_1, \dots, afl_k\}$ is the set of atomic fields, where: $\forall$ i $\in$ [1..k], an atomic field $afl_i \in AFL$ is defined as a pair $(N, V)$ where:
    - $afl_i.N$ is the name of $afl_i$,
    - $afl_i.V$ is the value of $afl_i$,
  - $CFL = \{cfl_1, \dots, cfl_l\}$ is the set of complex fields, where: $\forall$ j $\in$ [1..l], a complex field $cfl_j \in CFL$ is defined as a pair $(N, FL')$ where:
    - $cfl_j.N$ is the name of $cfl_j$,
    - $cfl_j.FL' \in FL$ is the set of fields that $cfl_j$ contains. These fields can be atomic or complex.

### 4.1.2 Delete query

A delete query $Q^D$ is defined as a pair $(Cll$, Id) where:

- $Q^D.Cll$ is the name of the collection on which $Q^D$ is executed,
- $Q^D.Id$ is the identifier of the document to delete.

### 4.1.3 Update query

An update query $Q^U$ is defined as a triple $(Cll$, Id, $FL)$ where:

- $Q^U.Cll$ is the collection name mentioned in $Q^U$,
- $Q^U.Id$ is the identifier of the document to update,
- $Q^U.FL = AFL \cup CFL$ is a set of fields to update in the document where the identifier is $Id$, where:
  - $AFL = \{afl_1, \dots, afl_k\}$ is a set of atomic fields, where : $\forall$ i $\in$ [1..k], an atomic field $afl_i \in AFL$ is defined by a couple $(N, V)$ where:
    - $afl_i.N$ is the field name,
    - $afl_i.V$ is the field value.

- $CFL = \{cfl_1, \ldots, cfl_l\}$ is a set of complex fields, where: $\forall$ j $\in$ [1..l], a complex field $cfl_j \in CFL$ is defined by a couple ($N$, $FL'$) where:

- $cfl_j.N$ is the field name,
- $cfl_j.FL' \in FL$ is a set of fields that compose $cfl_j$.

These fields can be either atomic or complex.

Query metamodel is shown in Figure2; this metamodel describes the main concepts used in insertion, update and delete queries.

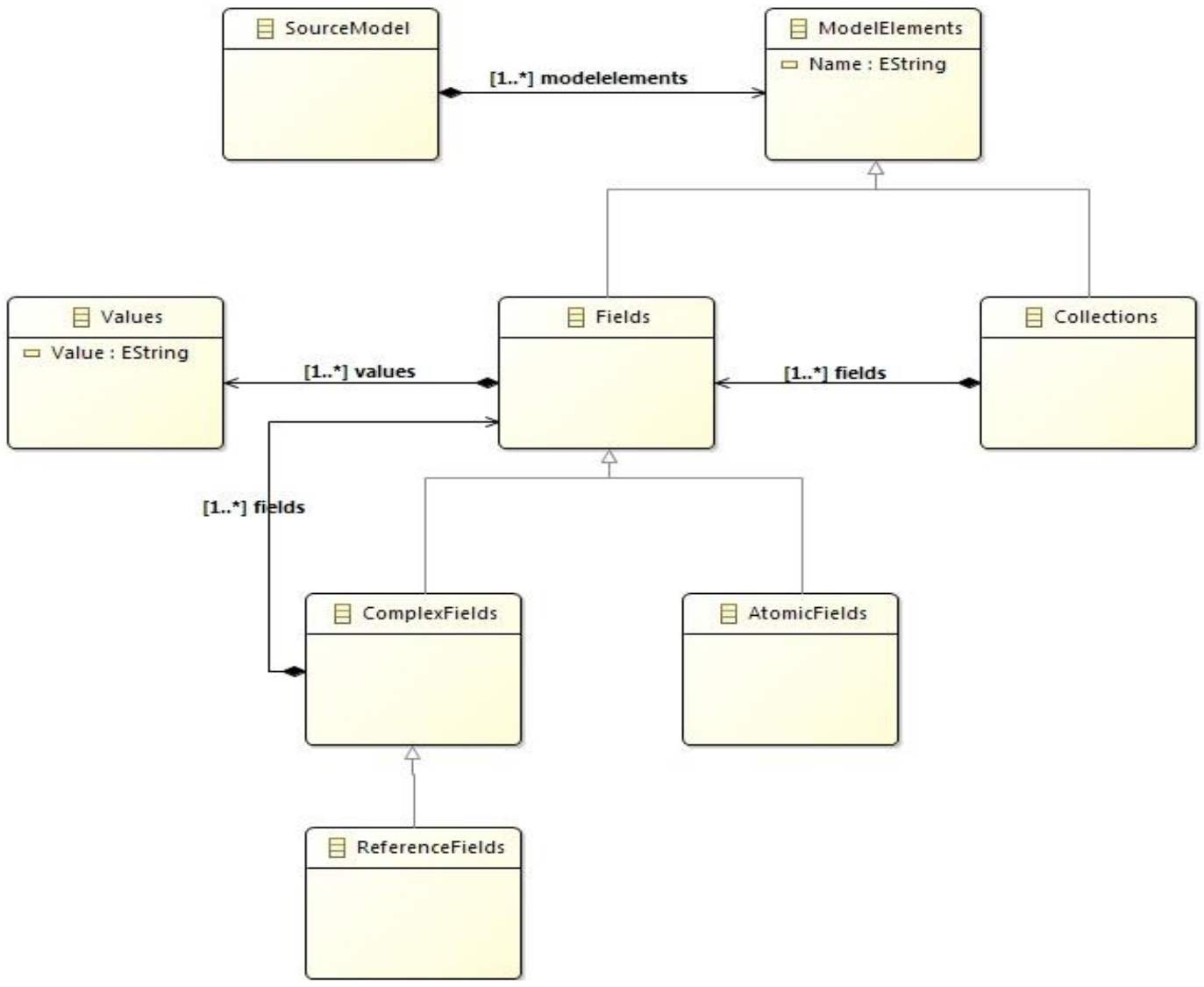

**Fig. 2.** Query metamodel

### 4.1.4 NoSQL model

The NoSQL model is stored in a JSON objet noted $O^M$. This is defined as a pair (Id, $CAT$), where:

- $O^M.Id$ is the identifier of the object $O^M$,
- $O^M.CAT = \{cat_1, \ldots, cat_n\}$ is the set of complex attributes, where: $\forall$ i $\in$ [1..n], a complex attribute $cat_i$ is defined as a pair ($N$, $AT$) where:

- $cat_i.N$ is the name of $cat_i$. It corresponds to the name of a collection in the database,

- $cat_i.AT = AAT \cup CAT'$ is a set of attributes of the collection, where:

- $AAT = \{aat_1, \ldots, aat_k\}$ is the set of atomic attributes of the collection, where: $\forall$ j $\in$ [1..k], an atomic attribute $aat_j$ is defined as a pair ($N$, $T$) where:

- $aat_j.N$ is the name of $aat_j$,
- $aat_j.T$ is the type of $aat_j$.

- $CAT' = \{cat_1', \ldots, cat_l'\}$ is the set of complex attributes of the collection, where: $\forall$ j $\in$ [1..l], a complex attribute $cat_j'$ is defined as a pair ($N$, $AT'$) where:

- $cat_j'.N$ is the name of $cat_j'$,
- $cat_j'.AT'$ is the set of attributes that $cat_j'$ contains. These attributes can be atomic or complex.

### 4.1.5 Metadata

Metadata is stored in a JSON objet noted $O^{MD}$. This is defined as a pair (Id, $CAT$), where:

- $O^{MD}.Id$ is the identifier of the object $O^{MD}$,
- $O^{MD}.CAT = \{cat_1, \ldots, cat_n\}$ is the set of complex attributes, where: $\forall$ i $\in$ [1..n], a complex attribute $cat_i$ is defined as a pair ($N$, $AT$) where:

- $cat_i.N$ is the name of $cat_i$. It corresponds to the name of a collection in the database,

- $cat_i.AT = AAT \cup CAT'$ is a set of attributes of the collection, where:

- $AAT = \{aat_1, \ldots, aat_k\}$ is the set of atomic attributes of the collection, where: $\forall$ j $\in$ [1..k], an atomic attribute $aat_j$ is defined as a pair ($N$, $Cpt$) where:

- $aat_j.N$ is the name of $aat_j$,
- $aat_j.Cpt$ is the number of occurrences of $aat_j$ in the collection where it appears,

- $CAT' = \{cat_1', \ldots, cat_l'\}$ is the set of complex attributes of the collection, where: $\forall$ j $\in$ [1..l], a complex attribute $cat_j'$ is defined as a pair ($N$, $AT'$) where:

- $cat_j'.N$ is the name of $cat_j'$,
- $cat_j'.AT'$ is the set of attributes that $cat_j'$ contains. These attributes can be atomic or complex.

We present the concepts used to describe the NoSQL model and the metadata through the meta-model of Figure 3.

## 4.2 Transformation rules

In this section, we present our Query2Model process as a sequence of transformation rules described below for each type of query.

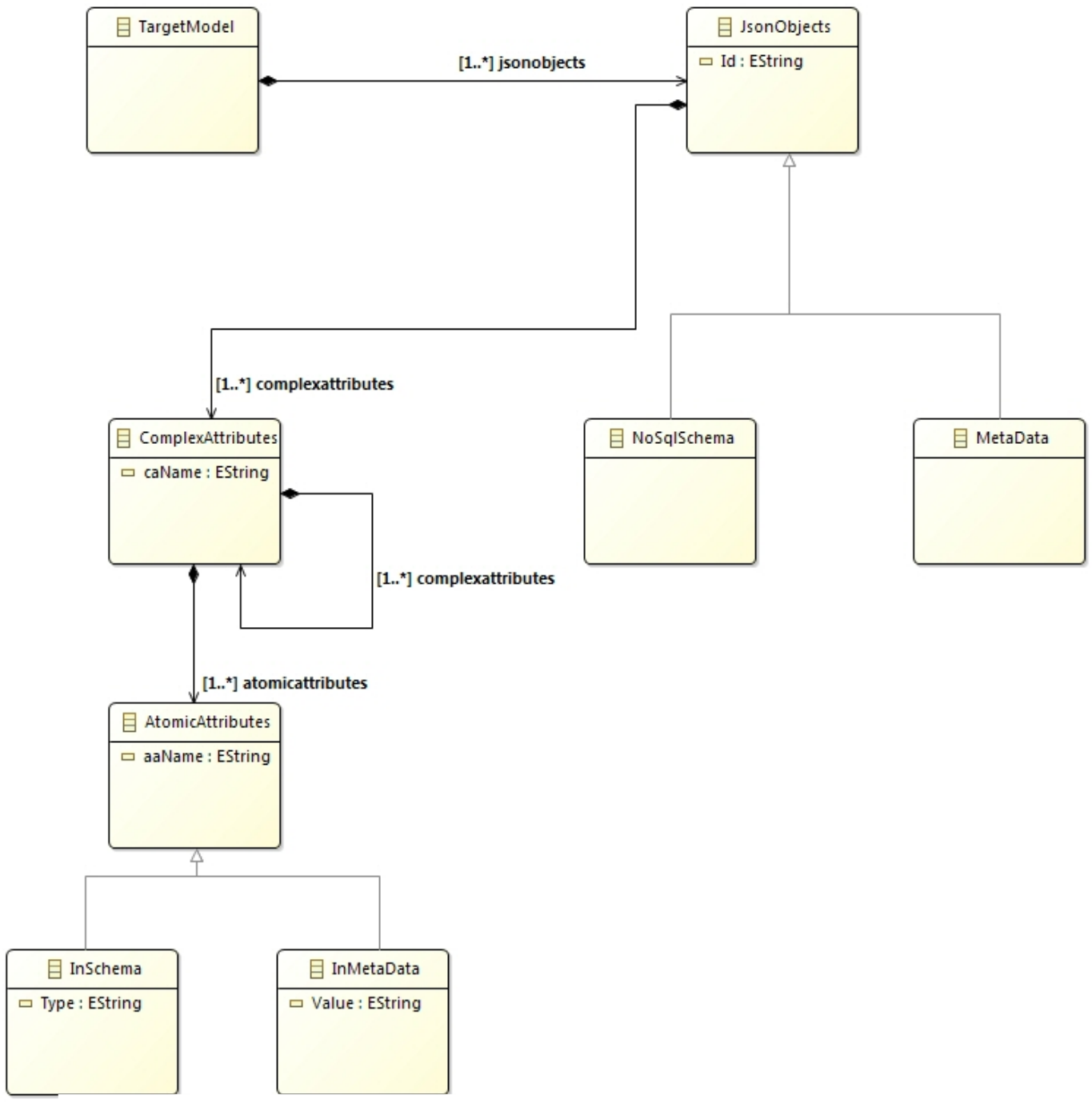

**Fig. 3.** Metadata and NoSQL metamodel

### 4.2.1 Case of an insertion query

Triggered by an insert query containing the name of the collection and a set of fields of the form (Name, Value), Query2Model process proceeds as follows:

**Step 1:** It compares the elements of the query with the current model of the collection and updates it, if necessary.
**Step 2:** It enriches metadata that will be used later for the treatment of delete query.

Formally, for each insertion query $Q^I$, Query2Model process applies rules R1 and R2:

**R1:** Each pair (Name, Value) in the insertion query is transformed into a pair (Name, Type) in the model as follows:

- If there is an attribute $cat_i \in O^M.CAT$ , where i ∈ [1..n], and $cat_i.N = Q^I.Cll$, then:
  - For each field $afl_j \in Q^I.AFL$ , with j ∈ [1..k] :
    - Transform the pair $(afl_j.N, afl_j.V)$ into a pair $(afl_j.N, afl_j.T)$.
    - If there is not an attribute $aat_j \in cat_i.AAT$, with i ∈ [1..n] and j ∈ [1..k], where: $aat_j.N = afl_j.N$ and $aat_j.T = afl_j.T$, then:
      - Create a pair $(aat_j.N, aat_j.T)$ and add it to $cat_i.AAT$, as:
        - $aat_j.N = afl_j.N$,
        - $aat_j.T$ is generated based on $afl_j$.
- Else:
  - Create an attribute $cat_i \in O^M.CAT$ , with i ∈ [1..n], as $cat_i.N = Q^I.Cll$
    - For each field $afl_j \in Q^I.AFL$ , with j ∈ [1..k] :
      - Create a pair $(aat_j.N, aat_j.T)$ and add it to $cat_i.AAT$, as:
        - $aat_j.N = afl_j.N$
        - $aat_j.T$ is generated based on $afl_j$.

**R2:** Each pair (Name, Value) in the insertion query is transformed into a pair (Name, Cpt) in the metadata as follows:

- If there is an attribute $cat_i \in O^{MD}.CAT$ , with i ∈ [1..n], as $cat_i.N = Q^I.Cll$, then:
  - For each field $afl_j \in Q^I.AFL$ , with j ∈ [1..k]:
    - Transform the pair $(afl_j.N, afl_j.V)$ into a pair $(afl_j.N, afl_j.Cpt)$.
    - If there is not an attribute $aat_j \in cat_i.AAT$, with i ∈ [1..n] and j [1..k], as: $aat_j.N = afl_j.N$ and $aat_j.T = afl_j.T$, then:
      - Create a pair $(aat_j.N, aat_j.Cpt)$ and add it to $cat_i.AAT$, as:
        - $aat_j.N = afl_j.N$
        - $aat_j.Cpt = 1$
    - Else,
      - Update the pair $(aat_j.N, aat_j.Cpt)$, as:
        - $aat_j.N = afl_j.N$
        - $aat_j.Cpt = aat_j.Cpt + 1$
- Else,
  - Create an attribute $cat_i \in O^{MD}.CAT$ , with i ∈ [1..n], as: $cat_i.N = Q^I.Cll$
    - For each field $afl_j \in Q^I.AFL$ , with j ∈ [1..k] :
      - Create a pair $(aat_j.N, aat_j.Cpt)$ and add it to $cat_i.AAT$, as:
        - $aat_j.N = afl_j.N$
        - $aat_j.Cpt = 1$.

### 4.2.2 Case of a delete query

A delete query contains the name of the collection and the identifier of the document to be deleted. Query2Model process treats this type of queries as follows:
**Step 1:** It recalculate the model based on the current metadata of the collection.
**Step 2:** It updates the metadata.
Formally, for each delete query $Q^D$, Query2Model process applies rules R3 and R4:

**R3:** Each field of the document whose identifier is Id is transformed as follows:

- Look for the attribute $cat_i \in O^{MD}.CAT$ , with i ∈ [1..n], where $cat_i.N = Q^D.Cll$:
  - Find the corresponding pair $(aat_j.N, aat_j.Cpt)$, as $aat_j \in cat_i.AAT$, with i ∈ [1..n] and j ∈ [1..k]
    - If $aat_j.Cpt = 1$, then:
      - Delete the corresponding pair $(aat_j.N, aat_j.T)$ from the model $O^M$.

**R4:** Each pair $(aat_j.N, aat_j.T)$ deleted from the model $O^M$, is transformed into a pair $(aat_j.N, aat_j.Cpt)$ in metadata $O^{MD}$

- Look for the attribute $cat_i \in O^{MD}.CAT$, with i ∈ [1..n], as $cat_i.N = Q^D.Cll$:
  - $aat_j.Cpt = aat_j.Cpt - 1$

### 4.2.3 Case of an update query

An update query contains the name of the collection, the identifier of the document to be modified and a set of fields of the form (Name, Value) to be updated in the document in question. The Query2Model process treats this type of queries as follows:

**Step 1:** It updates the model based on the current metadata of the collection
**Step 2:** It updates the metadata for future queries.

Formally, for each change request $Q^U$, the Query2Model process applies the following rules:

**R5:** If it is an insertion of a field in the document in question, then apply R1 and R2.

**R6:** If it is a deletion of a field in the document, then apply R3 and R4.

**R7:** If it is about renaming a field in the document, then:

- Look for $cat_i \in O^{MD}.CAT$ , with i $\in$[1..n], and $cat_i.N = Q^U.Cll$:

- Find the corresponding pair ($aat_j.N$, $aat_j.Cpt$), where $aat_j \in cat_i.AAT$, i $\in$ [1..n] and j $\in$ [1..k]

- If $aat_j.Cpt = 1$, then:

- Delete the corresponding pair ($aat_j.N$, $aat_j.T$) from the model $O^M$ and create another pair with the new attribute name.

- Delete the corresponding pair ($aat_j.N$, $aat_j.Cpt$) from the metadata $O^{MD}$ and create another pair with the new attribute name.

- Else:

- Keep the corresponding pair ($aat_j.N$, $aat_j.T$) in the model $O^M$ and create another with the new attribute name.

- Keep the corresponding pair ($aat_j.N$, $aat_j.Cpt$) in the metadata $O^{MD}$ and create another with the new attribute name.

**R8:** If it is about updating the value of a field in the document, then:

- Look for $cat_i \in O^M.CAT$ , with i $\in$[1..n], and $cat_i.N = Q^U.Cll$:

- Find the corresponding pair ($aat_j.N$, $aat_j.T$), where $aat_j \in cat_i.AAT$, i $\in$ [1..n] and j $\in$ [1..k]

- If the two values of the field are not of the same type, then:

- If $aat_j.Cpt = 1$, then:

- Delete the corresponding pair ($aat_j.N$, $aat_j.T$) in the model $O^M$and create another pair with the new attribute type.

- Else:

- Keep the corresponding pair ($aat_j.N$, $aat_j.T$) in the model $O^M$ and create another pair with the new attribute type.

- Keep the corresponding pair ($aat_j.N$, $aat_j.Cpt$) in the metadata $O^{MD}$ by decreasing $Cpt$ and create another pair with the new attribute type.

# 5 Experiments and validation

## 5.1 Experiments

To demonstrate the practical applicability of our work, we have implemented the process defined above in Eclipse Modeling Framework (EMF) that provides a convenient environment for formalizing transformations. It combines a set of several powerful modeling standards. Among these standards we used: (1) XML Metadata Interchange (XMI) for exchanging metadata information via XML and (2) Query / View / Transformation (QVT) language for specifying transformations. Transformations presented in section 4.2 are expressed as a sequence of elementary steps that builds the resulting NoSQL model step by step from the input query. First, we implement the transformation rules by means of the QVT plugin provided within EMF. Then, we test the transformation by running the QVT script. This script takes as input (1) the current database model version and (2) a query provided by the user and returns as output the NoSQL database model new version.

To illustrate this, we build an insert query using the standard-based XML Metadata Interchange (XMI) format. The QVT script is then executed on this query to update the complex field that contain the model of the collection provided in the insert query. The result after running the script is provided in the form of XMI file.

## 5.2 Validation

### 5.2.1 Experimental environment

Our problem is to extract the model of a database managed by a NoSQL system. Such a feature is intended for users who do not know the data structure (developer who has not created the database, decision makers, etc.); its major interest is to allow the expression of queries as can be done in relational systems. The experiments of our proposal were carried out on a cluster composed of 3 machines. Each machine has the following specifications: Intel Core i5, 8 GB of RAM and 2 TB of disk. One of these machines is configured to act as a master; the other two machines have slave status. To implement our solution, we used [17] AND [18] to generate data. We produced a 3TB dataset in the form of JSON files. These files were loaded into MongoDB using shell commands.

### 5.2.2 Query set

For our experiment, we have considered four kinds of queries: **(1)** those using one collection (example : select the patients whose age is between 10 and 70), **(2)** queries that use two related collections with the link is expressed using a monovalued reference field (example: we want the name of doctor who has performed the consultation number 41), **(3)** queries that use two related collections with the link is expressed using a multivalued reference field (example: select the antecedents of patient ”DUPONT David”), **(4)** queries that use two related collections with the link is expressed using nested data. Table 2 shows the comparison results between our solution and those proposed in [11], [12], [7], [14], [15] and [16] regarding the expression of queries. Note, however, that we only consider works that deal with document-oriented NoSQL databases. Thus, we have excluded the work of [13] which uses a graph-oriented database. For each query we have considered to perform this comparison, we indicate if it can be formulated using the model obtained by each solution proposed in the mentioned works.

| Query category | [11] | [12] | [7] | [14] | [15] | [16] | QueryToModel |
|---|---|---|---|---|---|---|---|
| (1) | X | X | X | X | X | X | X |
| (2) | | | | | | | X |
| (3) | | | | | | | X |
| (4) | | X | | | | | X |

Table. 2 Comparison results between our solution and state of the art

Table 2 shows that the absence of taking into account the links between collections in the referenced works [11], [12], [7], [14], [15] and [16], does not make it possible to write complex queries. Considering for example the following query that applies a join between the Patients collection and the Doctors collection:

**db.Patients.aggregate (**
**[**
**{$ lookup: {from: "Doctors", localField: "TreatingDoctors._id", foreignField: "_id", as: "Doctors"}}**
**])**

We can see that we can not write this query if we do not visualize the link between Patients and Doctors.

# 6 Conclusion and perspectives

Our work is part of the evolution of databases towards Big Data. They are currently focused on the extraction mechanisms of the model of a NoSQL database "schema less" to allow the expression of queries by end-users. In this article, we have proposed an automatic process that builds the physical model of a NoSQL database as it is used. This process is based on the Model Driven Architecture (MDA) architecture that provides a formal framework for automating model transformations. Our process generates a NoSQL physical model from insert, delete and update queries by applying a sequence of transformations formalized with the QVT standard. The returned model describes the structure of the collections that make up the database as well as the links between them. We have experimented our process on the case of a medical application that deals with scientific programs for the follow-up of pathologies; the database is stored on the MongoDB system.

Regarding future work, we aim to enrich our process so that it can take into consideration the diversity of particular cases related to the data entered. In fact, when feeding the database, users can enter incorrect data: misspelled field names, values associated with the same field of different types, etc. The current version of our process is based on consistent strategies, but the result may not be entirely satisfactory to users.

# 7 References

[1] Angadi, A. B., Angadi, A. B., & Gull, K. C. (2013). Growth of New Databases & Analysis of NOSQL Datastores. International Journal of Advanced Research in Computer Science and Software Engineering, 3, 1307-1319.

[2] Abdelhedi, F., Brahim, A. A., Atigui, F., & Zurfluh, G. (2017, August). MDA-Based Approach for NoSQL Databases Modelling. In International Conference on Big Data Analytics and Knowledge Discovery (pp. 88-102). Springer, Cham.

[3] Bondiombouy, C. (2015). Query processing in cloud multistore systems. In BDA : Bases de Données Avancées.

[4] Budinsky, F., Steinberg, D., Ellersick, R., Grose, T. J., & Merks, E. (2004). Eclipse modeling framework: a developer's guide. Addison-Wesley Professional.

[5] Chen, CL Philip et Zhang, Chun-Yang. Data-intensive applications, challenges, techniques and technologies: A survey on Big Data. Information Sciences, 2014, vol. 275, p. 314-347.

[6] Douglas, L., 2001. 3d data management: Controlling data volume, velocity and variety. Gartner. Retrieved, 6, 2001.

[7] Gallinucci, E., Golfarelli, M., & Rizzi, S. (2018). Schema profiling of document-oriented databases. Information Systems, 75, 13-25.

[8] Han, Jing, Haihong, E., LE, Guan, et al. Survey on NoSQL database. Pervasive computing and applications (ICPCA), 2011 6th international conference on. IEEE, 2011. p. 363-366.

[9] Harrison, G. (2015). Next Generation Databases : NoSQLand Big Data. Apress.

[10] Hutchinson, J., Rouncefield, M., & Whittle, J. (2011, May). Model-driven engineering practices in industry. In Proceedings of the 33rd International Conference on Software Engineering (pp. 633-642). ACM.

[11] Klettke, M., U. Störl, et S. Scherzinger (2015). Schema extraction and structural outlier detection for json-based nosql data stores. Datenbanksysteme für Business, Technologie und Web (BTW 2015).

[12] Sevilla, Diego Ruiz, Severino Feliciano Morales, and Jesús García Molina. "Inferring versioned schemas from NoSQL databases and its applications." International Conference on Conceptual Modeling. Springer, Cham, 2015.

[13] Comyn-Wattiau, I., & Akoka, J. (2017, December). Model driven reverse engineering of NoSQL property graph databases: The case of Neo4j. In 2017 IEEE International Conference on Big Data (Big Data) (pp. 453-458). IEEE.

[14] Maity, B., Acharya, A., Goto, T., & Sen, S. (2018, June). A Framework to Convert NoSQL to Relational Model. In Proceedings of the 6th ACM/ACIS International Conference on Applied Computing and Information Technology (pp. 1-6). ACM.

[15] Baazizi, M. A., Lahmar, H. B., Colazzo, D., Ghelli, G., & Sartiani, C. (2017, March). Schema inference for massive JSON datasets. In Extending Database Technology (EDBT).

[16] Baazizi, M. A., Colazzo, D., Ghelli, G., & Sartiani, C. (2019). Parametric schema inference for massive JSON datasets. The VLDB Journal, 1-25.

[17] Generate Test Data (2018) http://www.convertcsv.com/generate-test-data.htm Online; 5 July 2018.

[18] JSON Generator (2018) http://www.json-generator.com/. Online; 5 July 2018.